\documentclass[aps,twocolumn,showpacs,preprintnumbers,amsmath,amssymb,pra,unsortedaddress,superscriptaddress,longbibliography,letterpaper,floatfix]{revtex4-2}
\usepackage[latin9]{inputenc}
\usepackage{xcolor}
\usepackage{amsmath}
\usepackage{amssymb}
\usepackage{stmaryrd}
\usepackage{graphicx}
\usepackage[normalem]{ulem}
\usepackage{remreset} 
\usepackage{bm}
\usepackage[unicode=true,bookmarks=true,bookmarksnumbered=false,bookmarksopen=false,breaklinks=false,pdfborder={0 0 1},backref=false,colorlinks=true]
 {hyperref}
\hypersetup{
 linkcolor=magenta, urlcolor=blue, citecolor=blue, pdfstartview={FitH},  unicode=true}

\makeatletter
\usepackage{appendix}
\usepackage{amsfonts}
\usepackage{tabularx}
\usepackage{dcolumn}
\usepackage{bm}
\usepackage{graphicx}
\usepackage{epstopdf}
\usepackage{times}
\usepackage{lineno}
\hypersetup{urlcolor=blue}

\def\be{\begin{equation}}
\def\ee{\end{equation}}
\def\bea{\begin{eqnarray}}
\def\eea{\end{eqnarray}}
\def\bal{\begin{aligned}}
\def\eal{\end{aligned}}

\makeatother

\begin{document}
\nolinenumbers
\title{Grand Canonical-like Thermalization of Quantum Many-body Scars}

\author{Jia-wei Wang}
\affiliation{Anhui Province Key Laboratory of Quantum Network, University of Science and Technology of China, Hefei, 230026, China}
\affiliation{Synergetic Innovation Center of Quantum Information and Quantum Physics, University of Science and Technology of China, Hefei, 230026, China}
\affiliation{ Hefei National Laboratory, University of Science and Technology of China, Hefei 230088, China}

\author{Xiang-Fa Zhou}\email{xfzhou@ustc.edu.cn}
\affiliation{Anhui Province Key Laboratory of Quantum Network, University of Science and Technology of China, Hefei, 230026, China}
\affiliation{Synergetic Innovation Center of Quantum Information and Quantum Physics, University of Science and Technology of China, Hefei, 230026, China}
\affiliation{ Hefei National Laboratory, University of Science and Technology of China, Hefei 230088, China}
\affiliation{Anhui Center for Fundamental Sciences in Theoretical Physics, University of Science and Technology of China}

\author{Guang-Can Guo}
\affiliation{Anhui Province Key Laboratory of Quantum Network, University of Science and Technology of China, Hefei, 230026, China}
\affiliation{Synergetic Innovation Center of Quantum Information and Quantum Physics, University of Science and Technology of China, Hefei, 230026, China}
\affiliation{ Hefei National Laboratory, University of Science and Technology of China, Hefei 230088, China}

\author{Zheng-Wei Zhou}\email{zwzhou@ustc.edu.cn}
\affiliation{Anhui Province Key Laboratory of Quantum Network, University of Science and Technology of China, Hefei, 230026, China}
\affiliation{Synergetic Innovation Center of Quantum Information and Quantum Physics, University of Science and Technology of China, Hefei, 230026, China}
\affiliation{ Hefei National Laboratory, University of Science and Technology of China, Hefei 230088, China}
\affiliation{Anhui Center for Fundamental Sciences in Theoretical Physics, University of Science and Technology of China}

\begin{abstract}
Quantum many-body scar (QMBS) in kinetically constrained quantum systems challenges the conventional eigenstate thermalization hypothesis (ETH).
We develop an effective open-system description for constrained dynamics and introduce the definition of quasiparticle number in the system.
Based on this, we formulate a revised ETH that accounts for both diagonal and off-diagonal structures of local observables.
By introducing the cross coherence purity (CCP), we obtain a unified characterization of off-diagonal matrix elements and show that the relevant density of states (DOS) is determined by the distribution of eigenstates on the energy--quasiparticle-number plane.
We numerically verify an inverse relation between the CCP and this generalized DOS.
Applied to the quantum many-body scar model, the revised ETH accurately predicts long-time averages and temporal fluctuations of local observables and explains their dependence on initial states.
Our framework shows that the anomalous fluctuations and quasi-periodic dynamics of scar states arise naturally from low-DOS regions.
These results provide a unified understanding of thermalization and
QMBS in kinetically constrained systems.
\end{abstract}

\maketitle

\section{Introduction}

The eigenstate thermalization hypothesis (ETH) is a fundamental framework describing the thermalization process of isolated quantum many-body systems~\cite{eth1_1991,eth2_1994,eth3_2011,eth4_2016,eth5_2018}. This hypothesis provides an explanation at the eigenstates-level for why such systems reach thermal equilibrium during long-time evolutions~\cite{eth6_1996,eth7_1999,eth8_2008,eth10_2009,eth11_2010,eth12_2013,eth13_2014}. Specifically, it posits that the expectation values of any local observable $\hat{O}$ in the eigenstates $| E_i \rangle$, termed eigenstate expectation values (EEVs) $\langle E_i| \hat{O}|E_i\rangle$, approximates the micro-canonical average at the corresponding eigenenergies $E_i$, denoted as $\mathcal{O}(\mathcal{E})$. Additionally, the off-diagonal matrix elements of the observable between different eigenstates, $\langle E_i| \hat{O}|E_{i'}\rangle$ for $i \neq i'$, are sufficiently small, specifically bounded by the inverse square root of the density of states: $\Omega(\mathcal{E})^{-1/2}$. Under such assumptions, it follows that for an experimentally realizable initial state $|\psi(0)\rangle$, the long-time average of any observable's expectation value approaches the functional value $\mathcal{O}(\mathcal{E})$ at the initial energy ${\cal E}=\langle \psi(0)| H|\psi(0) \rangle$, which also converges to canonical average in the thermodynamic limits. Furthermore, the temporal fluctuations of this expectation value are bound by the inverse square root of DOS, thus rendering them relatively small.

The above logic is generally applicable to all non-integrable models. However, exceptions do exist. A notable instance is the quantum many-body scars (QMBS)~\cite{scarforall,quantumscarexp_2017,fractionpxp,highspinPXP,cjturner1,Yinglei1,Yinglei2,Yinglei3}, which violate various aspects of the ETH. This phenomenon is typically marked by the emergence of exceptional eigenstates known as scar states. While most eigenstates thermalize, a non-thermal subspace of scar eigenstates persists, scaling extensively with system size~\cite{scartower,SMembedding1,SMembedding2,SGA1,spin1kitaevmodel,QMBSsyzh}.
These states exhibit EEVs that significantly deviate from the canonical average and are separated by approximately uniform energy level spacings~\cite{zhaihui,pxpsga,systematic1d}.
These features lead to non-thermal phenomenon under certain initial states---The long-time averages of local observables no longer approach the canonical averages of the system, with pronounced quasi-periodic oscillation over time, indicating that the time-fluctuation cannot be bound by the DOS. 
Particularly, several kinetically constrained models, such as the well-known PXP model, have been shown to host QMBS. The reasons behind these violations remain unclear, raising fundamental questions about the potential thermalization of such systems.

In our previous work~\cite{GCT_QMBS}, we demonstrate that blockade-induced kinetic constraints can be realized by coupling unconstrained dynamics to an auxiliary environment. We employ two distinct dissipation channels representing information exchanges between the system and the environment.
Consequently, we find that the thermalization in such systems may encompass not only energy but also the implementation of the kinetic constraints.
Within the ETH framework, we refine the hypothesis regarding the EEVs of local observable, suggesting that these values could be unified by a two variables function $\mathcal{O}({\cal E,N})$ of not only the energy ${\cal E}$  but also the quasi-particle number ${\cal N}$, which is employed to describe the equilibrium of information exchange between the system and the auxiliary environment. Correspondingly, we demonstrate that the eigenstates in such constrained systems can be regarded as the equilibrium states of a grand canonical-like ensemble. As a result, we show that under any realizable initial states, the long-time average of any local observable would converge to a grand canonical-like average, rather than the traditional canonical average.

In this paper, we further develop the viewpoint that thermalization in kinetically constrained models is jointly governed by energy and quasiparticle number, and should therefore be formulated within a grand-canonical framework consistent with the ETH. From this perspective, the conventional DOS $\Omega({\cal E})$ must be generalized to a two-variable function $\Omega({\cal E,N})$ defined on the energy-quasiparticle-number plane, where richer spectral structures emerge.
Within this framework, scar states are naturally interpreted as eigenstates residing in regions of anomalously low DOS. The large temporal fluctuations observed for certain initial states are likewise attributed to the sparse distribution of eigenstates in the ${\cal E-N}$ plane. Furthermore, the pronounced temporal fluctuations observed for certain initial states can be attributed to the sparse distribution of eigenstates in these regions. Finally, the quasi-periodic dynamics of quantum scars has traditionally been attributed to an approximate algebraic structure among the scar states, known as the spectrum-generating algebra (SGA). Our results show that such algebraic structures can naturally emerge in the regions of low DOS.

In Section II, we introduce a kinetically constrained model on a one-dimensional (1D) spin chain and describe its dynamics using a Lindblad-like equation from an open system perspective.
In Section III, motivated by this framework, we modify the diagonal terms of the ETH by parametrizing the EEVs of local observables in terms of both energy and quasiparticle number. We also provide numerical evidence that long-time averages of observables can be well captured by the corresponding grand canonical-like ensemble.
In Section IV, we further revise the off-diagonal ETH by generalizing the concept of DOS to energy-quasiparticle-number plane. Moreover, We define a cross coherence norm (CCN) to quantify the off-diagonal contributions and show that the generalized DOS bounds both the amplitudes of the off-diagonal matrix elements of local observables and the temporal fluctuations in the long-time dynamics.
In Section V, we clarify the conditions for the emergence of SGA in scarred systems and demonstrate that these conditions can be satisfied in the low DOS regime.
We summarize this work in the final Section.

\section{One-dimensional Constrained Spin Models}

As a concrete platform for analytical derivations and numerical simulations, we study a one-dimensional (1D) spin chain of length $D$ with periodic boundary condition(PBC)~\cite{quantumscarexp_2017,QMBS1Dspin,GCT_QMBS}. In the absence of constraints, the dynamics is generated by the free Hamiltonian $H_0 = \sum_{k=1}^{D} s_k^x$, where $s_k^x$ denotes the spin operator along the $x$-direction at the $k$-th site. Each site hosts a spin of magnitude $j$, with $j \geq 1$.

Kinetic constraints are imposed by enforcing blockades that prohibiting adjacent spins from occupying the product states $|x\rangle_{k,k+1} = |j\rangle_k \otimes |-j\rangle_{k+1},\ k=1,2,\cdots,D$.
The dynamics is therefore restricted to a constrained Hilbert subspace $\mathcal{H}$, defined by the projector $\hat{P} = \prod_{k=1}^{D} \left( \mathbb{I} - |x\rangle_{k,k+1}\langle x| \right)$.
The constrained dynamics is typically obtained by projecting the free Hamiltonian onto $\mathcal{H}$: $\hat{P} H_0 \hat{P} = H \hat{P} = \hat{P} H$, where $H$ commutes with $\hat{P}$ and explicitly reads
\begin{equation}\label{HamiltonianH}
\begin{aligned}
H &= \sum_{k=1}^{D} s_k^x - (s_k^x + s_{k+1}^x) |x\rangle\langle x|_{k,k+1} - \mathrm{h.c.} \\
  &= \sum_{k=1}^{D} s_k^x - \sqrt{j}\, M_k - \sqrt{j}\, M_k^{\dagger}.
\end{aligned}
\end{equation}
Here, $M_k = \frac{1}{\sqrt{j}} |x\rangle\langle x|_{k,k+1} (s_k^x + s_{k+1}^x) = |x\rangle\langle y|_{k,k+1}$ map the states $|y\rangle_{k,k+1} = \frac{1}{\sqrt{j}} (s_k^x + s_{k+1}^x) |x\rangle_{k,k+1}$ within $\mathcal{H}$ to the blockaded states $|x\rangle_{k,k+1}$.
The last two terms in Eq.(\ref{HamiltonianH}) encode the kinetic constraints: the term $-\sqrt{j} M_k$ suppresses transitions from $\mathcal{H}$ to the forbidden configurations $|x\rangle_{k,k+1}$, while its Hermitian conjugate suppresses the reverse processes.

This model exhibits quantum many-body scarring. In particular, under the initial state $|\psi_1(0)\rangle = |j,j,\dots,j\rangle$, the system shows pronounced quasi-periodic revivals during the time evolution. Besides, for the special case $j=1$, this model can be mapped onto the PXP model defined on a chain of doubled length~\cite{QMBS1Dspin,GCT_QMBS,scarprojection}.

\begin{figure}
	\centering
	\includegraphics[width=8.5cm]{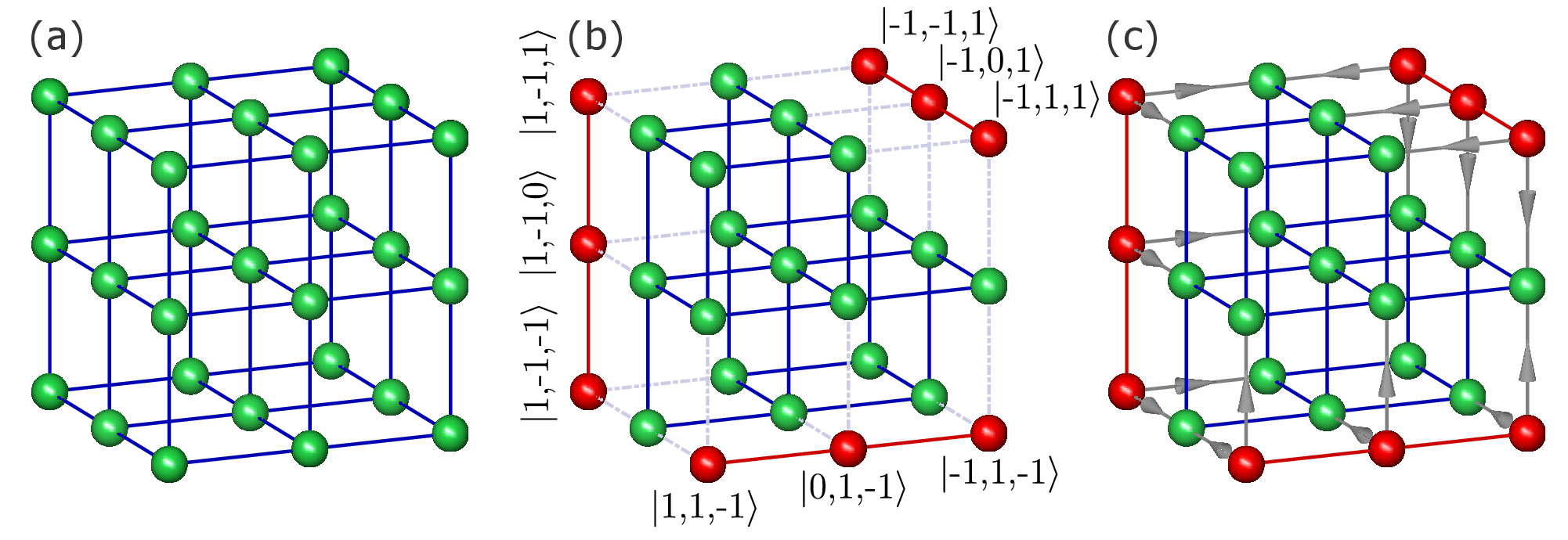}
	\caption{ Adjacency graphs in Hilbert space demonstrating the actions of the Hamiltonian. We consider a three-site chain with spin size $j=1$, where nodes represent all product states and lines indicate the mappings between them. Plot (a) shows the basic actions generated by the free Hamiltonian $H_0$, The action of $H_0$ forms a hypercubic structure in the adjacency graph, with each line denoting a bidirectional mapping between the two endpoint states. Plot (b) depicts the constrained Hamiltonian $H$, in which all the states containing patterns as $|1\rangle_k\otimes |-1\rangle_{k+1}(k=1,2,3)$ are blockaded, with light gray lines indicating the prohibited mappings. Plot (c) illustrates the effect of the non-Hermitian Hamiltonian $H_N$. Arrowed lines indicate unidirectional mappings from the prohibited states to the constrained subspace, corresponding to the terms $M_k^{\dagger}$ in Eq.(\ref{nonHermitianHamiltonian}). In plots (b) and (c), the green nodes represent the states within the constrained subspace ${\cal H}$, while the red ones denote prohibited states. As seen in plot (c) that, when the system is initiated within ${\cal H}$, the $M_k^{\dagger}$ terms will not affect the constrained dynamics.}
	\label{fig: Demonstration}
\end{figure}

A natural question arises: can the kinetic constraint be equivalently described by dissipative mechanism in an open-system framework?
We notice that, when restricting to dynamics within the constrained subspace $\mathcal{H}$, the term $M_k^{\dagger}$ in Eq.~\eqref{HamiltonianH} does not contribute, since $M_k^{\dagger} \hat{P}= |y\rangle\langle x|_{k,k+1} \hat{P}= 0$.
Thus, for dynamics initiated in $\mathcal{H}$, it suffices to only suppress the transitions from $\mathcal{H}$ to the forbidden subspace. This allows us to replace the original constraint by an anti-Hermitian term without modifying the effective dynamics within $\mathcal{H}$, As illustrated in the schematic diagrams of Fig.~\ref{fig: Demonstration}. We therefore introduce the non-Hermitian Hamiltonian
\begin{equation}
H_N
= \sum_{k=1}^{D} s_k^x
  - \sqrt{j} M_k
  + \sqrt{j} M_k^{\dagger}
  - i c \sqrt{\tfrac{j}{2}}\, \pi_k ,
\label{nonHermitianHamiltonian}
\end{equation}
where $\pi_k = |x\rangle\langle x|_{k,k+1}$ and $c>0$ is a tunable real parameter. The last two terms leave the constrained dynamics invariant due to $\hat{P}|x\rangle_{k,k+1}=0$. Together with $-\sqrt{j} M_k$, these three terms can be engineered via dissipative mechanism.
Specifically, we introduce the dissipation channels:
\begin{equation}
\left\{
\begin{aligned}
& \gamma_1 = c \sqrt{2j},
  & L_{k,1} = \pi_k - \frac{i\sqrt{2}}{c} M_k , \\
& \gamma_2 = -\sqrt{2j}\,\frac{2}{c},
  & L_{k,2} = M_k .
\end{aligned}
\right.
\label{DissipationChannels}
\end{equation}
Combined with the free Hamiltonian $H_0$, these channels define a Lindblad-like equation:
\begin{equation}
\partial_t \rho
= \mathcal{L}(\rho)
\equiv -i H_N \rho + i \rho H_N^{\dagger} + \mathcal{J}(\rho),
\label{LindbladlikeEq}
\end{equation}
where $H_N = H_0 - \frac{i}{2} \sum_{\sigma,k} \gamma_\sigma L_{k,\sigma}^{\dagger} L_{k,\sigma}$ explicitly is the Hamiltonian from Eq.(\ref{nonHermitianHamiltonian}). The jumping term is given by \begin{equation}\label{JumpingTerms}
  \begin{aligned}
  \mathcal{J}(\rho) = & \sum_{k,\sigma} \gamma_\sigma L_{k,\sigma} \rho L_{k,\sigma}^{\dagger} \\
    = &\sqrt{2j} \sum_{k=1}^{D} c \pi_k \rho \pi_k -i \sqrt{2} M_k \rho \pi_k + i \sqrt{2} \pi_k \rho M_k^{\dagger}.
  \end{aligned}
\end{equation}
Since $\pi_k \hat{P} = \hat{P} \pi_k = 0$, the jumping term annihilates all states within $\mathcal{H}$. Therefore, the non-Hermitian Hamiltonian $H_N$ generates the equivalent dynamics as the constrained Hamiltonian $H$ within $\mathcal{H}$, where the jumping term $\mathcal{J}$ also has no effect. The Liouvillian superoperator $\mathcal{L}$ therefore provides an equivalent open-system description of the kinetically constrained dynamics.

From a thermodynamic perspective, these dissipations implicitly introduce an auxiliary environment coupled to the system. The presence of negative dissipation rates $\gamma_2$ indicates non-Markovian dynamics and information backflow from the environment. 
However, since the dynamics governed by $\mathcal{L}$ is equivalent to the unitary dynamics $H$, the Lindblad-like equation remains positive definite and trace-preserving within ${\cal H}$~\cite{LindbladMasterEquation1,LindbladMasterEquation2,LindbladMasterEquation3,LindbladMasterEquation4,Zeno1,Zeno2}.
In this picture, the kinetic constraint emerges as a balance between information loss and recovery, with the information exchange rate governed by the jump probabilities~\cite{nonMarkovJump,QTD,QTD1,QTD2,QTD3}.
For any state $\rho(t)$ within $\mathcal{H}$, the probabilities of the two kinds of jumps are equal:
\begin{equation}
\begin{aligned}
P_+ &= \gamma_1 \sum_{k=1}^{D} \mathrm{Tr}(L_{k,1} \rho L_{k,1}^{\dagger}) \\
&= \frac{2\sqrt{2j}}{c} \sum_{k=1}^{D}\mathrm{Tr}(M_k \rho M_k^{\dagger}) = P_-.
\end{aligned}
\label{JumpingRate}
\end{equation}
Hence, the information exchange rate is proportional to $\mathrm{Tr}(\rho \hat{N})$, where $\hat{N} = \sum_{k=1}^{D} M_k^{\dagger} M_k = \sum_{k=1}^{D} |y\rangle\langle y|_{k,k+1}$ is the counter of patterns $|y\rangle_{k,k+1}$.
The operator $\hat{N}$ can therefore be regarded as the quasiparticle number operator and $N_k=|y\rangle\langle y|_{k,k+1}$ as the local quasiparticles.
Previous study has shown that the scar states in this model exhibit significantly lower information exchange rates than nearby thermal states, and correspondingly anomalously low EEVs of $\hat{N}$~\cite{GCT_QMBS}.
These observations motivate the conjecture that, in kinetically constrained systems, the quasiparticle number $\hat{N}$, together with the energy, jointly characterizes the potential thermalization process and the resulting equilibrium.
More generally, the equivalence between kinetic constraints and the engineered dissipative dynamics, as well as the associated thermodynamic perspective based on open system, applies beyond the specific model studied here. This formulation was systematically introduced and discussed in our earlier work~\cite{GCT_QMBS}.

\begin{figure}
	\centering
	\includegraphics[width=8.5cm]{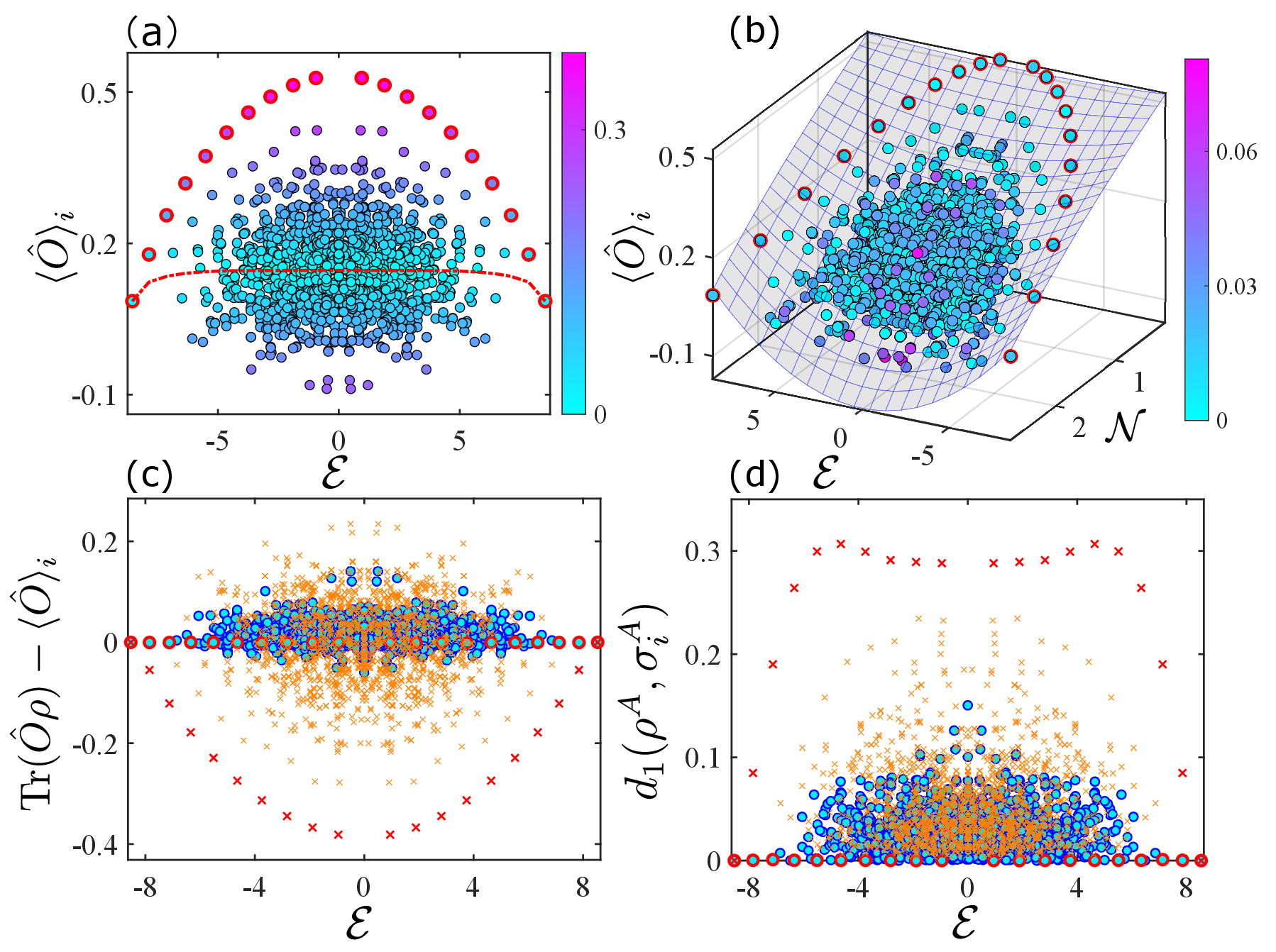}
	\caption{Plot (a) shows the EEVs of a local observable $\hat{O}=s_1^z \otimes s_2^z$, with $s_k^z$ being the $z$-direction spin operator on $k$-th site. The red line represents a single-parameter fit $\mathcal{O}(\mathcal{E})$ as a function of the energy $\mathcal{E}$. Plot (b) displays the same EEVs fitted as a smooth function of both $\mathcal{E}$ and the quasi-particle number $\mathcal{N}$. In plots (a) and (b), scar eigenstates are highlighted by red circles, while the color bars indicate the deviation of each eigenstate from the fitted curve (or surface). These results show that the EEVs cannot be described solely by the energy, with scar states exhibiting particularly large deviations, whereas all states are unified by a smooth function of $\mathcal{E}$ and $\mathcal{N}$. Plot (c) shows the difference between the EEVs and the ensemble averages: $\mathrm{Tr}(\hat{O}\rho)-\langle \hat{O}\rangle_i$, where $\rho$ is taken as either the canonical ensemble $\rho_c(\beta_c)$ or the grand canonical-like ensemble $\rho(\beta,\mu)$. Plot (d) shows the trace-norm distance between the reduced DMs of eigenstates, $\sigma_i^A=\mathrm{Tr}_{\bar{A}}(|E_i\rangle\langle E_i|)$, and that of the corresponding ensembles, $\rho^A=\mathrm{Tr}_{\bar{A}}(\rho)$, with the local region $A$ being a single site. In plots (c) and (d), crosses denote the canonical ensemble results, while dots correspond to the grand canonical ensemble. Results for scar states are marked in red. It can be seen that the grand canonical ensemble provides significantly more accurate predictions for the local properties of eigenstates, particularly for scar states. We employ a spin chain with length $D=10$ and spin size $j=1$, This setting is used throughout the main text. 
}\label{fig: Diagonal_terms}
\end{figure}

\section{Eigenstates as Grand Canonical-like Equilibrium States}

The ETH provides a universal description of thermalization in generic non-integrable quantum many-body systems~\cite{eth7_1999}. For a local observable $\hat{O}$, ETH postulates that its matrix elements in the eigenbasis of the Hamiltonian take the form:
\begin{equation}
  \langle E_i |\hat{O} |E_{i'}\rangle
  = \mathcal{O}(\mathcal{E}) \delta_{i,i'}
  + \Omega(\mathcal{E})^{-1/2} f(\mathcal{E},\omega) R_{i,i'},
  \label{ETH}
\end{equation}
where $|E_i\rangle$ denotes an eigenstate with eigenenergy $E_i$, $\mathcal{E} = (E_i + E_{i'})/2$ and $\omega = E_i - E_{i'}$ are the mean energy and energy difference, respectively. The functions $\mathcal{O}(\mathcal{E})$ and $f(\mathcal{E},\omega)$ are smooth functions of these arguments, $\Omega(\mathcal{E})$ is the DOS at energy $\mathcal{E}$, and $R_{i,i'}$ is a pseudo-random variable with zero mean and unit variance.

Separating Eq.~(\ref{ETH}) into its diagonal and off-diagonal components, the diagonal part implies that each eigenstate can be regarded as a microcanonical equilibrium state. 
Consequently, the EEVs of a local observable depend smoothly on the energy,
$\langle E_i|\hat{O}|E_i\rangle \simeq \mathcal{O}(E_i)$.
For a general initial state $|\psi(0)\rangle = \sum_{i=1}^{\mathcal{D}} c_i |E_i\rangle$, the long-time average of the observable is given by
\begin{equation}
\begin{aligned}
  \bar{O}
  &\equiv \lim_{T \rightarrow \infty} \frac{1}{T} \int_0^T
  \langle \psi(t)| \hat{O} |\psi(t)\rangle \, dt \\
  &= \sum_{i=1}^{\mathcal{D}} |c_i|^2
  \langle E_i | \hat{O} | E_i \rangle \\
  &= \mathcal{O}(\mathcal{E})
  + \mathbb{O}(\Delta_E^2)
  + \mathbb{O}(\Omega^{-1/2}),
\end{aligned}
\label{longtimeaverageETH}
\end{equation}
where $\mathcal{E} = \sum_{i=1}^{\mathcal{D}} |c_i|^2 E_i$ is the mean energy of the initial state and $\Delta_E^2 = \sum_i (E_i - \mathcal{E})^2 |c_i|^2$ is its energy variance. Eq.(\ref{longtimeaverageETH}) assumes a sufficiently dense and non-degenerate spectrum, as well as a small energy uncertainty $\Delta_E$.
Finally, the function $\mathcal{O}(\mathcal{E})$ is commonly identified with the canonical ensemble average: $\mathcal{O}(\mathcal{E}) \simeq \mathrm{Tr}[\rho_c(\beta_c)\hat{O}]$, where $\rho_c(\beta_c) = e^{-\beta_c H}/Z$ is the canonical DM. The inverse temperature $\beta_c$ is determined by the fixing the energy: $\mathcal{E} = \mathrm{Tr}[\rho_c(\beta_c) H]$~\cite{eth7_1999}.

In the kinetically constrained models introduced above, a subset of scar eigenstates exhibits clear violations of the ETH. In particular, the EEVs of local observables for these states cannot be described as smooth functions of energy alone, as demonstrated in Fig.~\ref{fig: Diagonal_terms}.

Motivated by the open-system thermodynamic interpretation developed in Sec.~II, we assume that each eigenstate $|E_i\rangle$ can still be regarded as an equilibrium state, but that such equilibrium is characterized not only by the energy $E_i$ but also by the rate of information exchange with the auxiliary environment. This additional degree of freedom is naturally quantified by the quasiparticle number,
$N_i = \langle E_i | \hat{N} | E_i \rangle$.
We therefore propose a revised diagonal ETH of the form
\begin{equation}
  \langle E_i | \hat{O} | E_i \rangle
  \simeq \mathcal{O}(\mathcal{E},\mathcal{N}),
  \label{rETHdiagonal}
\end{equation}
where $\mathcal{E} = E_i$ and $\mathcal{N} = N_i\equiv \langle E_i| \hat{N} | E_i\rangle$ denote the energy and quasi-particle number of the eigenstate, respectively, and $\mathcal{O}(\mathcal{E},\mathcal{N})$ is a smooth function of both variables.

Following the same reasoning that leads to Eq.~(\ref{longtimeaverageETH}), the long-time average of a local observable for an initial state $|\psi(0)\rangle = \sum_i c_i |E_i\rangle$ becomes
\begin{equation}
\begin{aligned}
  \bar{O}
  &= \sum_{i=1}^{\mathcal{D}} |c_i|^2
  \langle E_i | \hat{O} | E_i \rangle \\
  &\simeq \mathcal{O}(\mathcal{E},\mathcal{N})
  + \mathbb{O}(\Delta_E^2)
  + \mathbb{O}(\Delta_N^2),
\end{aligned}
\label{longtimeaverageRETH}
\end{equation}
where $\mathcal{E} = \sum_i |c_i|^2 E_i$ and $\mathcal{N} = \sum_i |c_i|^2 N_i$ are the long-time averaged energy and quasi-particle number, respectively. The quantity $\Delta_N^2 = \sum_i (N_i - \mathcal{N})^2 |c_i|^2$ denotes the variance of the quasi-particle number. Equation~(\ref{longtimeaverageRETH}) holds provided that both $\Delta_E$ and $\Delta_N$ are sufficiently small.

This revised diagonal ETH naturally motivates a grand canonical-like ensemble description,
\begin{equation}
  \rho(\beta,\mu)
  = \frac{\sum_{i=1}^{\mathcal{D}}
  e^{-\beta (E_i - \mu N_i)}
  |E_i\rangle\langle E_i|}
  {\sum_{i=1}^{\mathcal{D}} e^{-\beta (E_i - \mu N_i)}},
  \label{GrandCanonicalEnsemble}
\end{equation}
where $\beta$ and $\mu$ play the roles of inverse temperature and chemical potential, respectively. $\mathcal{D}$ marks the dimension of the constrained subspace ${\cal H}$.
The function $\mathcal{O}(\mathcal{E},\mathcal{N})$ can then be approximated by the ensemble average $\mathrm{Tr}[\hat{O}\rho(\beta,\mu)]$, with $\beta$ and $\mu$ determined by fixing the energy and the quasi-particle number:
$\mathcal{E} = \mathrm{Tr}[\rho(\beta,\mu) H]$ and
$\mathcal{N} = \mathrm{Tr}[\rho(\beta,\mu) \hat{N}]$.

The validity of the revised diagonal ETH in kinetically constrained systems is supported by several numerical observations. Firstly, as shown in Fig.~\ref{fig: Diagonal_terms} (a) and (b), the EEVs of local observables cannot be collapsed onto a smooth function of energy alone, but can be accurately described by a smooth two-variable function of energy $\mathcal{E}$ and quasi-particle number $\mathcal{N}$. Within this unified description, both thermal eigenstates and scar eigenstates are captured by the same function $\mathcal{O}(\mathcal{E},\mathcal{N})$.
Secondly, Fig.~\ref{fig: Diagonal_terms} (c) and (d) demonstrate that the EEVs of all eigenstates---including scar states---are well approximated by the grand canonical-like ensemble average defined in Eq.~(\ref{GrandCanonicalEnsemble}). In contrast, the predictions of the conventional canonical ensemble show clear and systematic deviations for scar eigenstates. Notably, the grand canonical ensemble provides particularly accurate predictions for the EEVs of scar states.
In addition, we compute the trace-norm distance between the reduced density matrices of individual eigenstates and those of the two ensembles, in order to eliminate possible biases arising from the specific choice of observables~\cite{Schattenp,AsignT_Eigen}. The resulting data further confirm the above conclusions.
Thirdly, for a physically realizable initial state $|\psi(0)\rangle$, the real-time dynamics of the expectations of local observables, shown in Fig.~\ref{fig: Timefluctuations} (a)-(c), reveals that while the long-time average deviates from the canonical ensemble prediction, it converges closely to the value predicted by the grand canonical-like ensemble. This provides direct dynamical evidence supporting the revised diagonal ETH in the constrained systems.

\begin{figure}
	\centering
	\includegraphics[width=8.5cm]{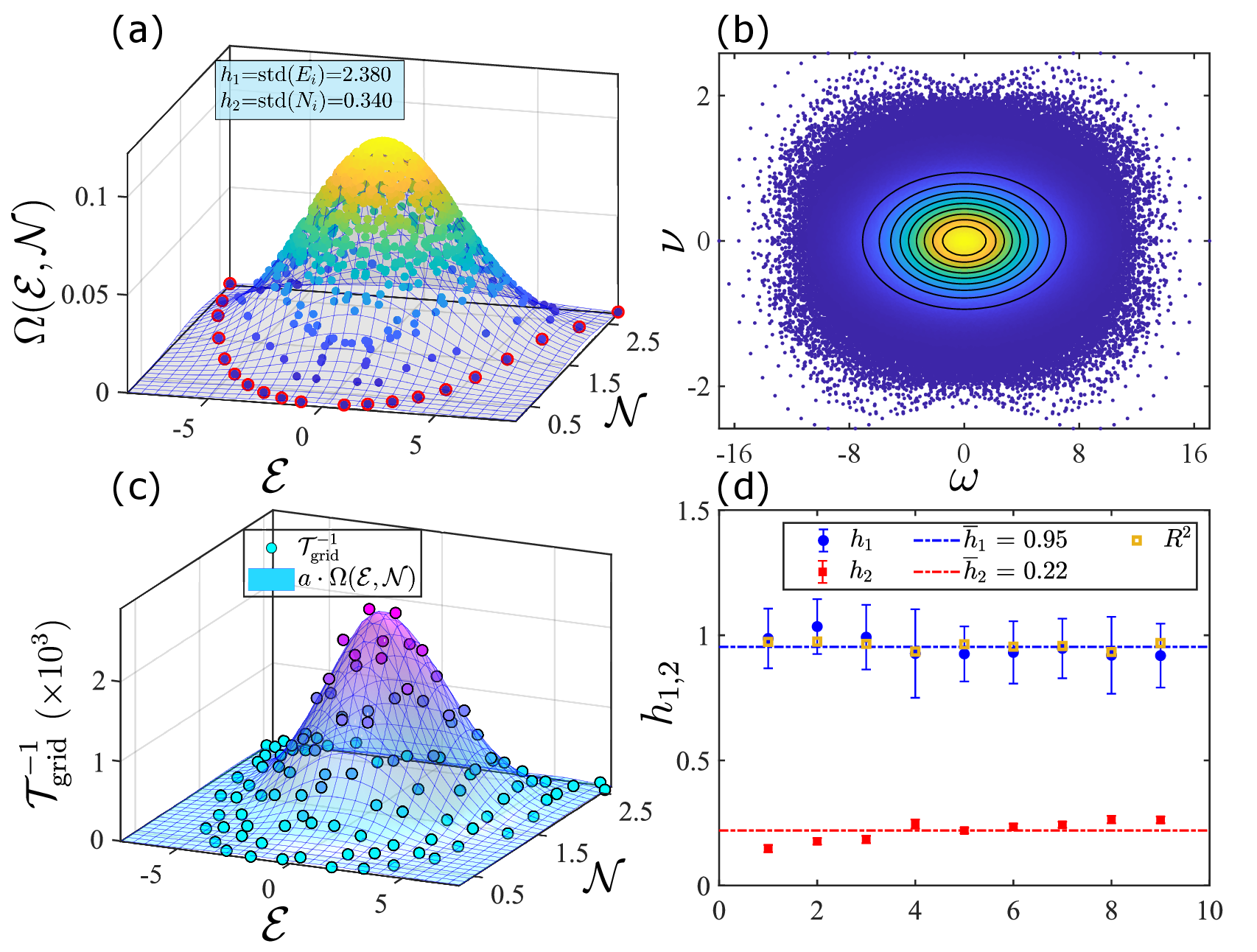}
	\caption{Plot (a) shows the DOS $\Omega(\mathcal{E},\mathcal{N})$ defined in Eq.~(\ref{DensityofStates}) on the energy--quasiparticle-number plane. The kernel widths $h_{1,2}$ are chosen as the standard deviations of the sets $\{E_i\}$ and $\{N_i\}$, respectively. The DOS values associated with all eigenstates are indicated by points on the surface, with scar states highlighted by red circles, revealing a pronounced suppression of DOS in the scarred region. Plot (b) shows the distribution of all off-diagonal terms on the $\omega$--$\nu$ plane, where the color encodes the local density of data points. Concentric ellipses enclosing $10\%, 20\%, \cdots, 90\%$ of the data are constructed from equal-density contours, and the enclosed subsets are used for statistical analysis. Plot (c) uses the innermost $10\%$ of the data points in the $\omega$--$\nu$ plane as samples. The $\mathcal{E}$--$\mathcal{N}$ plane is discretized into bins, within which the averaged quantities $\{E_{\mathrm{grid}}, N_{\mathrm{grid}}, \mathcal{T}_{\mathrm{grid}}\}$ are evaluated. The inverse averaged CCP, $\mathcal{T}_{\mathrm{grid}}^{-1}$, is then fitted by the function $a\,\Omega(E_{\mathrm{bin}},N_{\mathrm{bin}})$, showing excellent agreement with the generalized DOS. Plot (d) shows that, by varying the sample size from $10\%$ to $90\%$, different estimates of the kernel widths $h_{1,2}$ are obtained. The fitted values of $h_{1,2}$ for different sample percentages are shown as blue and red dots, with error bars indicating the corresponding standard deviations. The averaged kernel widths are shown as solid lines, while the coefficient of determination $R^2$, plotted as yellow points, remains high for all fittings.}
\label{fig: NonDiagonal_terms}
\end{figure}

\section{Time Fluctuations and the Density of States on the \({\cal E-N}\) plane}

According to the ETH, the off-diagonal matrix elements of a local observable $\hat{O}$ in a non-integrable system take the form $\langle E_i | \hat{O} | E_{i'} \rangle \big|_{i\neq i'}
= \Omega(\mathcal{E})^{-1/2} f(\mathcal{E},\omega)\, R_{i,i'}$. The off-diagonal elements should determine the temporal fluctuations of local observables under unitary evolution. For an initial state $|\psi(0)\rangle=\sum_i c_i |E_i\rangle$, the long-time averaged fluctuations are given by
\begin{equation}
\label{longtimeflucETH}
\begin{aligned}
\overline{(O(t)-\bar{O})^2}
&= \lim_{T\to\infty}\frac{1}{T}\int_0^T
\big(\langle \psi(t)|\hat{O}|\psi(t)\rangle-\bar{O}\big)^2\,dt \\
&= \sum_{i\neq i'} |c_i|^2 |c_{i'}|^2 |O_{i,i'}|^2 \\
&\sim \mathbb{O}\!\left(\Omega^{-1}\right).
\end{aligned}
\end{equation}
Since the DOS typically scales with the Hilbert-space dimension $\mathcal{D}$, which tends to infinity in the thermodynamic limit, conventional ETH predicts vanishing temporal fluctuations at long-time limit~\cite{eth2_1994,eth7_1999,eth8_2008}.

\begin{figure}
	\centering
	\includegraphics[width=8.5cm]{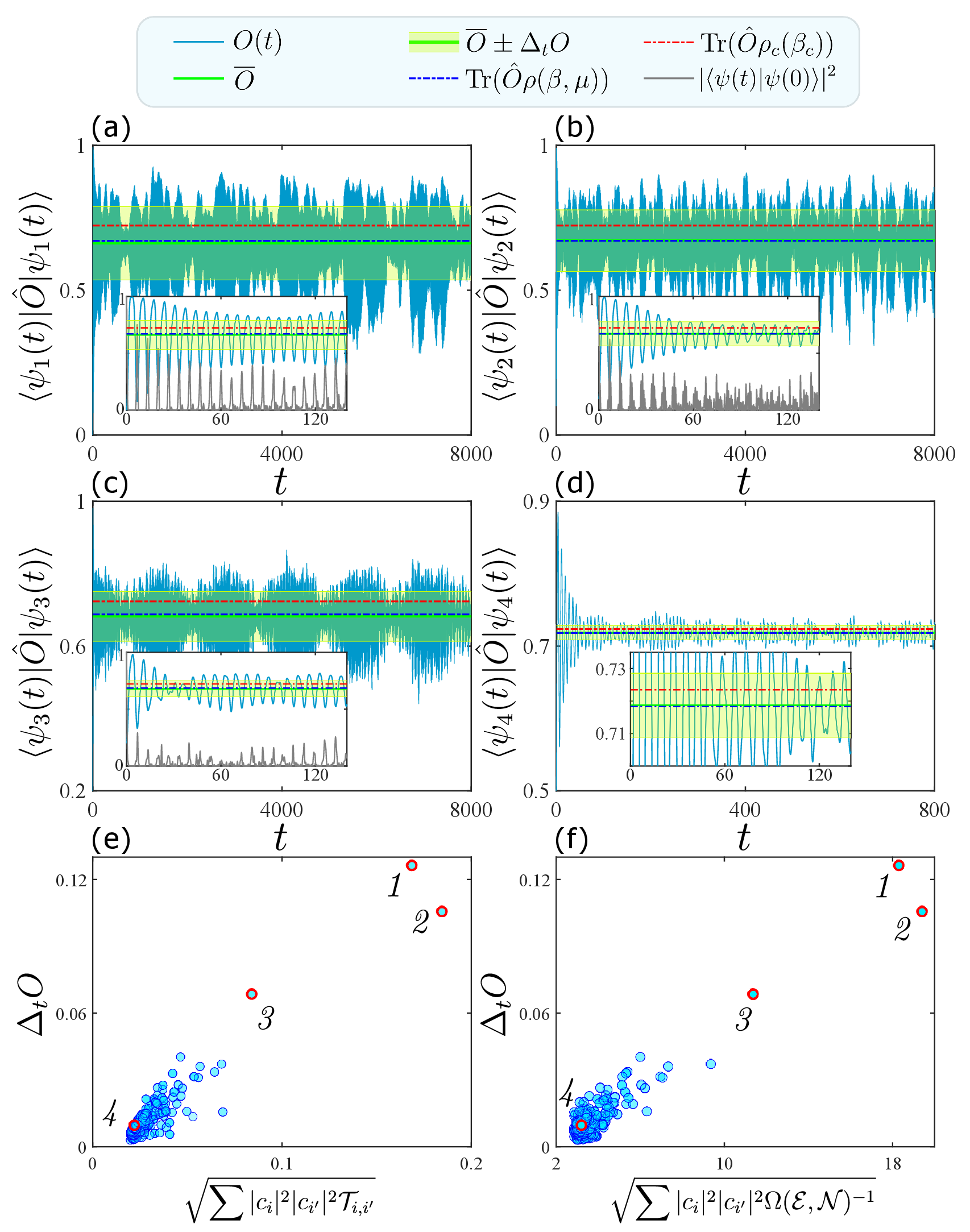}
	\caption{Plots (a)--(d) show the time evolution of a local observable $\hat{O}=|0\rangle\langle 0|_{1}+|-1\rangle\langle -1|_{1}$ for different initial states. 
The blue curves represent the expectation value $\langle \psi(t)|\hat{O}|\psi(t)\rangle$, while the green horizontal lines indicate its long-time average $\overline{O}$.
The dark-blue and red dotted lines denote the ensemble averages predicted by the grand canonical and canonical ensembles, respectively. 
The light-yellow shaded region shows the long-time averaged temporal fluctuations around the mean, $\overline{O}\pm\Delta_t O$. Insets display the short-time dynamics. 
In the insets of (a)--(c), the gray curves also show the fidelity evolution $|\langle \psi(t)|\psi(0)\rangle|^2$. We can see that the long-time average of the local observable is accurately captured by the grand canonical ensemble, indicating thermalization to a grand-canonical-like local state. 
In plots (e) and (f), all translation-invariant initial states are considered, and the long-time averaged temporal fluctuations are computed in the same manner as in (a)--(d). These results are compared with the estimates based on the CCP and the DOS, given by $\sqrt{\sum_{i\neq i'} |c_i|^2 |c_{i'}|^2\mathcal{T}_{i,i'}}$ and $\sqrt{\sum_{i\neq i'} |c_i|^2 |c_{i'}|^2\Omega(\mathcal{E},\mathcal{N})^{-1}}$, respectively. 
The observed temporal fluctuations are found to scale proportionally with these estimates.
}
\label{fig: Timefluctuations}
\end{figure}

In stark contrast to this hypothesis, the QMBS model studied here exhibits pronounced and persistent temporal oscillations for specific initial states. In particular, the fully polarized state $|\psi_1(0)\rangle = |j,j,\dots,j\rangle$ shows long-lived quasi-periodic revivals. More generally, initial states with a large overlap with scarred eigenstates, such as $|\psi_2(0)\rangle = \frac{1}{\sqrt{D}}\sum_{k=1}^{D}|1,1,\dots,0_k,\dots,1\rangle $, also exhibit anomalously large temporal fluctuations. In contrast, generic initial states display only small fluctuations consistent with the ETH, as demonstrated in Fig.~\ref{fig: Timefluctuations} (a)-(d) and Fig.~\ref{fig: SGA} (a) and (b).

These observations indicate that the suppression of the off-diagonal matrix elements by the energy-resolved DOS $\Omega(\mathcal{E})$ is insufficient to capture the dynamics of kinetically constrained systems.
Guided by the open-system interpretation and the modified diagonal ETH discussed above, we conjecture that the relevant DOS in this model is defined not solely along the energy axis, but on the joint energy--quasiparticle-number plane. We therefore propose the following modified ETH ansatz for the off-diagonal matrix elements of local observables:
\begin{equation}
\label{rETHnon-diagonal}
\langle E_i | \hat{O} | E_{i'} \rangle \big|_{i\neq i'}
= \Omega(\mathcal{E},\mathcal{N})^{-1/2}
\, f(\mathcal{E},\mathcal{N},\omega,\nu)\, R_{i,i'} .
\end{equation}
Here $\mathcal{E}=\frac{E_i+E_{i'}}{2}$ and $\omega=E_i-E_{i'}$ is inherited from the original ETH, while $\mathcal{N}=\frac{N_i+N_{i'}}{2}$ and $\nu=N_i-N_{i'}$ denote the mean quasiparticle number and the quasiparticle number difference, respectively. $\Omega(\mathcal{E},\mathcal{N})$ is the newly defined DOS on the ${\cal E-N}$ plane. $f$ is a smooth function in all its arguments, while $R_{i,i'}$ is a pseudo-random variable.

The low density of states $\Omega(\mathcal{E},\mathcal{N})$ in the vicinity of scar eigenstates, as shown in Fig.~\ref{fig: NonDiagonal_terms} (a), provides a natural explanation for the enhanced temporal fluctuations observed for scarred initial states.

In Eq.~\eqref{rETHnon-diagonal}, the function $f$ depends on the specific choice of the local observable, obscuring a direct verification of the DOS scaling. To eliminate this operator dependence, we introduce an operator-independent quantity that bounds off-diagonal matrix elements for all the local observables.
Given a local observable $\hat{O}=\sum_l O_l |l\rangle\langle l|$ supported on a subsystem $A$, one finds
\begin{equation}
\label{nondiagonal_matrix}
\begin{aligned}
\langle E_i | \hat{O} | E_{i'} \rangle
&= \mathrm{Tr}_A \!\left( \rho_A^{i',i} \hat{O} \right), \\
\rho_A^{i,i'} &= \mathrm{Tr}_{\bar{A}} \left(|E_i\rangle\langle E_{i'}|\right),
\end{aligned}
\end{equation}
where we name $\rho_A^{i,i'}$ the reduced cross-density matrix on subsystem $A$. Using the Cauchy--Schwarz inequality, one obtains the bound
\begin{equation}
\label{limitingNonDiagonal}
|\langle E_i | \hat{O} | E_{i'} \rangle|
\le
\max_l |O_l|\,
\sqrt{\mathcal{A}}\,
\sqrt{\mathrm{Tr}\!\left(
\rho_A^{i,i'\dagger}\rho_A^{i,i'}
\right)},
\end{equation}
with $\mathcal{A}$ the Hilbert-space dimension of subsystem $A$.
We define $\mathcal{T}_{i,i'} \equiv \mathrm{Tr}\!( \rho_A^{i,i'\dagger}\rho_A^{i,i'})$, and refer to it as the cross coherence purity (CCP). This quantity provides an operator-independent measure of the typical magnitude of off-diagonal matrix elements. Therefore, the modified ETH from Eq.(\ref{rETHnon-diagonal}) predicts $\mathcal{T}_{i,i'} \propto \Omega(\mathcal{E},\mathcal{N})^{-1}$. We note that the inequality in Eq.(\ref{limitingNonDiagonal}) is rigorous. However, it should be interpreted as an upper bound rather than an asymptotic equality, and therefore does not imply an exact scaling law.

For finite systems, eigenstates are discretely distributed on the $\mathcal{E}$--$\mathcal{N}$ plane. To obtain a smooth DOS, we introduce a Gaussian kernel estimator~\cite{DOSest},
\begin{equation}
\label{DensityofStates}
\Omega(\mathcal{E},\mathcal{N})
= Z \sum_{i=1}^{\mathcal{D}}\exp\!\left[ -\frac{(\mathcal{E}-E_i)^2}{2h_1^2} -\frac{(\mathcal{N}-N_i)^2}{2h_2^2}\right],
\end{equation}
where $h_{1,2}$ are kernel widths alongside the energy and quasiparticle-number directions, $Z=\frac{1}{2\pi \mathcal{D} h_1 h_2}$ is the normalization factor. $h_{1,2}$ are typically chosen as the standard deviations of the energy and quasiparticle number, as shown in Fig.~\ref{fig: NonDiagonal_terms}.
We compute CCP $\mathcal{T}_{i,i'}$ for all eigenstate pairs within fixed windows of $(\omega,\nu)$, bin the data on the $\mathcal{E}$--$\mathcal{N}$ plane, and extract averaged values
$\{\mathcal{E}_{\mathrm{grid}},\mathcal{N}_{\mathrm{grid}},\mathcal{T}_{\mathrm{grid}}\}$. As shown in Fig.~\ref{fig: NonDiagonal_terms} (a)-(c), the averaged inverse CCP closely follows the DOS $\Omega(\mathcal{E},\mathcal{N})$. By fitting $\mathcal{T}_{\mathrm{grid}}^{-1}$ with $a\cdot \Omega(\mathcal{E},\mathcal{N})$, we extract optimal kernel widths $h_{1,2}$, which remain stable under variations of the $(\omega,\nu)$ windows.

The modified ETH immediately yields a prediction for long-time temporal fluctuations,
\begin{equation}\label{longtimeflucRETH}
\begin{aligned}
\Delta_t^2 O \equiv \overline{(O(t)-\bar{O})^2}
&= \sum_{i\neq i'} |c_i|^2 |c_{i'}|^2 |O_{i,i'}|^2 \\
&\propto
\sum_{i\neq i'} |c_i|^2 |c_{i'}|^2
\, \Omega(\mathcal{E},\mathcal{N})^{-1},
\end{aligned}
\end{equation}
where we used the fact that $R_{i,i'}^2$ has unit mean. For a fixed local observable, the temporal fluctuations are therefore governed roughly by the DOS on the $\mathcal{E}$--$\mathcal{N}$ plane.

This prediction is confirmed numerically in Fig.~\ref{fig: Timefluctuations} (e) and (f), where all translation-invariant states, including the scarred initial states, are systematically used as initial conditions.

\section{Emergent Spectrum-Generating Algebra in the Low-Density Regime}

Building on the previous sections, the modified ETH framework not only accurately predicts the EEVs of local observables, but also provides reliable predictions for long-time dynamical properties, including both the expectation values and the temporal fluctuations of local observables. In contrast to the conventional ETH, the thermalization theory proposed for kinetically constrained models allows for distinct thermal behaviors in different regions of the energy--quasiparticle-number plane.

\begin{figure}
	\centering
	\includegraphics[width=8.5cm]{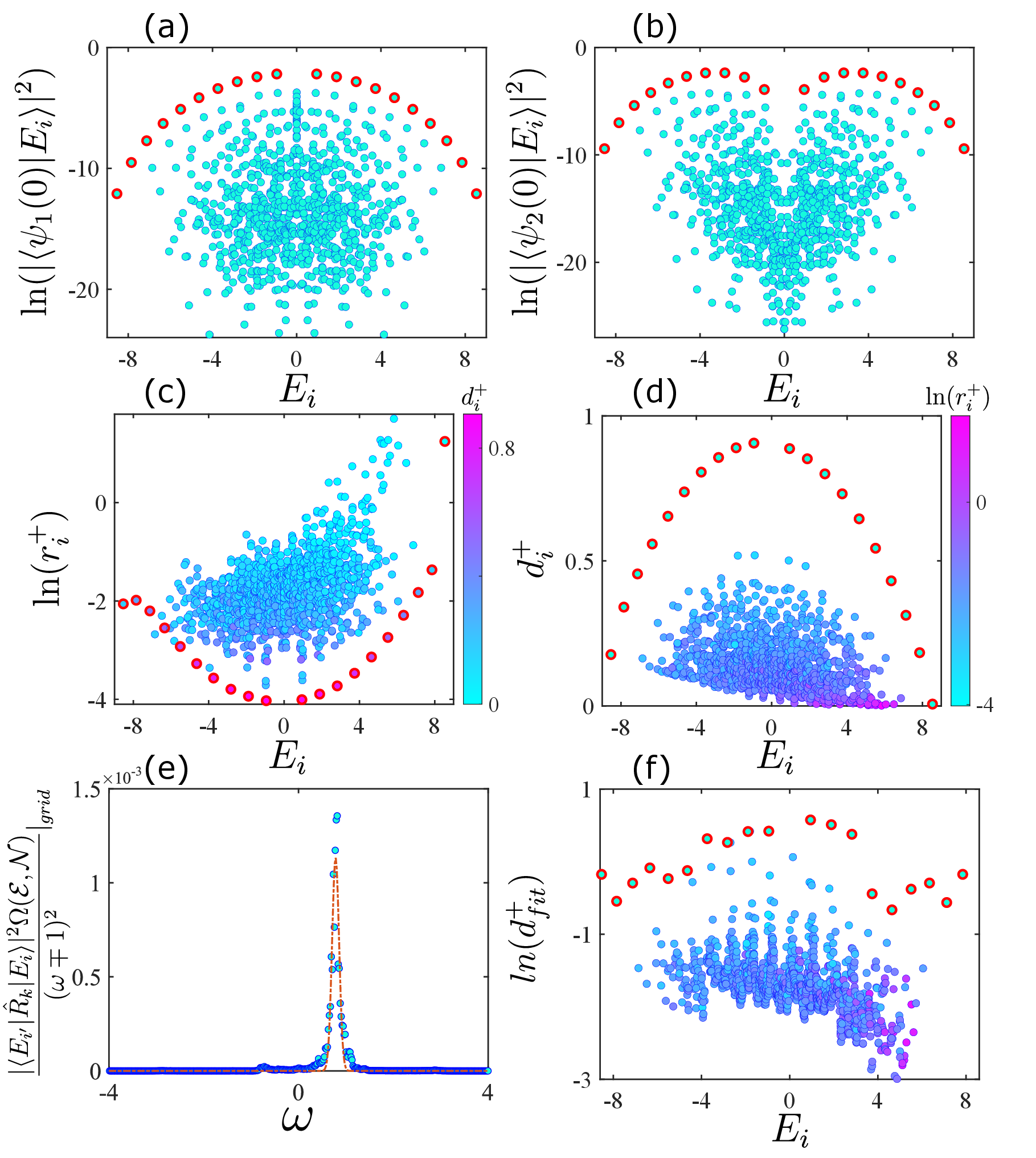}
	\caption{Plot (a) and (b) show the overlaps between eigenstates and two initial states, $|\psi_1(0)\rangle$ and $|\psi_2(0)\rangle$, which give rise to quasi-periodic dynamical oscillations, as defined in Sec.~IV. Scar eigenstates are highlighted by red circles. 
Plot (c) shows the ratio $r_i^{+}=n_i/d_i^{+}$ evaluated for all eigenstates, where the color bar indicates the magnitude of the denominator $d_i^{+}$. 
Plot (d) displays the denominator $d_i^{+}$ itself, with the color encoding the corresponding values of $r_i^{+}$. These results demonstrate that $r_i^{+}$ is essentially inversely correlated with $d_i^{+}$. 
Plot (e) shows that, after removing the DOS dependence, the squared off-diagonal matrix elements of the local observable $\hat{R}_k$, together with the factor $1/(\omega+1)^2$, form a single-peaked wave packet along the $\omega$ axis. The blue dots depict the averaged results after resolving the $\omega$ axis into bins. The red dotted line fits these dots with a gaussian function $g(\omega)$. 
By inserting the fitted $g(\omega)$ into the expression for $d_i^{+}$, we obtain the estimate $d_{\mathrm{fit}}^{+}$ predicted by the modified ETH, as shown in plot (f), where the color of each point represents the corresponding exact value $\ln(r_i^{+})$. 
In plots (a)-(d) and (f), scar eigenstates are marked by red circles.}
\label{fig: SGA}
\end{figure}

In the constrained models, quantum scars are characterized not only by anomalously large temporal fluctuations, but also by their nearly periodic dynamics under certain initial states. 
Previous studies have shown that this behavior can be traced back to the scar eigenstates, which show large overlaps with the chosen initial states. As shown in Fig.~\ref{fig: SGA}(a) and (b), these states form an approximately equally spaced energy ladder, and are linked by an approximate spectrum-generating algebra (SGA)~\cite{SGA1,pxpsga,QMBS1Dspin}.
A natural question then arises: why does such an algebraic structure emerge among scar states, but not among thermal eigenstates? 

For the present model, this structure is captured by the Lie-algebra-like relation~\cite{QMBS1Dspin}:
\begin{equation}
  [H,Q^{\pm}] = \pm Q^{\pm} + \hat{R}.
  \label{SGA1}
\end{equation}
where the operators $Q^{\pm}$ act as approximate ladder operators between adjacent scar states. The residual term $\hat{R}=i\frac{j}{\sqrt{2}}\sum_ {k=1}^{D}\hat{R}_k$, with $\hat{R} _k=|j-1,-j\rangle \langle j-1,-j|_{k,k+1}-|j,-j+1\rangle \langle j,-j+1|_{k,k+1}$, is a sum of local operators and therefore remains equivalent to the local terms $R_k$ within the translation-invariant subspace. 
Defining normalized states $|E_i^{\pm}\rangle \propto Q^{\pm}|E_i\rangle$, Eq.~(\ref{SGA1}) implies
\begin{equation}\label{SGA2}
  H |E_i^{\pm}\rangle = (E_i \pm 1)|E_i^{\pm}\rangle + \sqrt{r_i^{\pm}}\,|r\rangle,
\end{equation}
where the deviation from an exact ladder structure is quantified by:
\begin{equation}
\begin{aligned}
  r_i^{\pm} &= \frac{n_i}{d_i^{\pm}},\qquad n_i = \sum_{i'\neq i} |\langle E_{i'}|\hat{R}_k|E_i\rangle|^2, \\
  d_i^{\pm} &= \sum_{i'\neq i}
  \frac{|\langle E_{i'}|\hat{R}_k|E_i\rangle|^2}{(\omega \mp 1)^2},
\end{aligned}
\label{ReRefine_r}
\end{equation}
with $\omega = E_{i'} - E_{i}$. The ratio $r_i^{\pm}$ measures how accurately the state $|E_i^{\pm}\rangle$ approximates an eigenstate of $H$ with energy $E_i \pm 1$: smaller values of $r_i^{\pm}$ indicate a more precise equally spaced energy structure.

As shown in Fig.~\ref{fig: SGA} (c), scar eigenstates exhibit significantly smaller values of $r_i^{\pm}$ compared to thermal eigenstates, confirming the existence of an approximate SGA within the scar subspace. Importantly, Fig.~\ref{fig: SGA}(d) reveals that this reduction originates from an enhancement of the denominator $d_i^{\pm}$, while the numerator $n_i$ shows no obvious distinction between scar and thermal states (see Appendix).
This demonstrates that the approximate equally spaced structure of scar eigenstates is primarily controlled by the denominator in Eq.~(\ref{ReRefine_r}).

The magnitude of $d_i^{\pm}$ is governed by the off-diagonal terms of the local observable $R_k$. Besides, since $\hat{R}_k$ is local, its off-diagonal matrix elements should scale as $\Omega(\mathcal{E},\mathcal{N})^{-1/2}$, as predicted by the modified ETH and numerically verified in the Appendix.

Next, we explicitly extract the DOS-independent structure of the off-diagonal matrix elements. To this end, we consider the quantity $|\langle E_{i'}|\hat{R}_k|E_i\rangle|^2 \,\Omega(\mathcal{E},\mathcal{N})/(\omega\mp 1)^2$ as the full statistical sample.
According to the modified ETH, this object corresponds to $f(\mathcal{E},\mathcal{N},\omega,\nu)^2/(\omega\mp 1)^2$ multiplied by a random factor $R_{ii'}^2$. To suppress the random fluctuations, we focus on the dependence on the single variable $\omega=E_{i'}-E_i$. Specifically, we bin the data along the $\omega$ axis and average over each bin, which effectively eliminates the random factor $R_{i,i'}^2$. 
This procedure should yield a smooth function $g(\omega)$, capturing the systematic $\omega$ dependence of the DOS-independent part. In Fig.~\ref{fig: SGA} (e), this smooth function is simply obtained by fitting the data with a Gaussian function of $\omega$.
Finally, inserting the fitted function $g(\omega)$ into the expression for the denominator, we obtain an ETH-based estimate:
\begin{equation}\label{fit_d}
\begin{aligned}
  d_i^{\pm} &= \sum_{i'\neq i}
  \frac{|\langle E_{i'}|\hat{R}_k|E_i\rangle|^2}{(\omega \mp 1)^2} \\
  & = \sum_{i'\neq i}\frac{f(\mathcal{E},\mathcal{N},\omega,\nu)^2 \Omega(\mathcal{E},\mathcal{N})^{-1} R_{i,i'}^2 }{(\omega \mp 1)^2} \\
  & \approx \sum_{i'\neq i} \frac{g(\omega)}{\Omega(\mathcal{E},\mathcal{N})}=d_{\mathrm{fit}}^{\pm},
\end{aligned}
\end{equation}
which provides a quantitative approximation to $d_i^{\pm}$ for each
eigenstate. 

In Fig.~\ref{fig: SGA}(f), we present the ETH-based estimate $d_{\mathrm{fit}}^{+}$
for all eigenstates.
Distinctly larger $d_{\mathrm{fit}}^{+}$ is observed for the scar
eigenstates compared to the thermal ones.
This behavior roughly reproduces the structure found in the exact
denominator $d_i^{+}$.
More importantly, this result shows that within the revised ETH framework,
the emergence of a quasi-equally spaced energy ladder is a natural consequence
of the suppressed DOS in the corresponding
$\mathcal{E}-\mathcal{N}$ region.
In this sense, the approximate SGA observed among scar eigenstates is not an independent anomaly, but rather a nature outcome that is allowed in the modified thermalization theory.

\section{Conclusion}

In this work, we employed an effective open-system description for kinetically constrained models by constructing dissipative processes that capture the information exchanges underlying the constrained dynamics.
This perspective provides a unified framework to complete the thermalization theory for such systems and to analyze their long-time dynamical properties.
Within this framework, we established a revised ETH tailored to kinetically constrained systems, modifying both its diagonal and off-diagonal structures.
In particular, by introducing the CCP, we obtained a unified characterization of the off-diagonal matrix elements of local observables.
We showed that the relevant density of states from the traditional ETH is no longer determined solely by energy, but by the distribution of eigenstates on the energy--quasiparticle-number plane, and numerically verified an inverse relation between the CCP and this generalized DOS.
As a result, the revised ETH accurately predicts not only the long-time averages of local observables, but also the magnitude of their temporal fluctuations and their dependence on the initial state.
Applying this framework to the QMBS embedded in the constrained models, we demonstrated that the typically large fluctuations and quasi-periodic dynamics of the QMBS arise naturally from the low-DOS regions that in which the initial state locates.
In particular, the emergence of an approximate SGA and a quasi-equally spaced energy structure is shown to be compatible with ETH within the constrained setting, rather than a violation of it.
Our results provide a coherent understanding of thermalization in kinetically constrained systems and offer a broader perspective on QMBS.
More generally, this work sheds new light on the structure of ETH and on the
mechanisms governing non-equilibrium dynamics in isolated quantum many-body
systems.

\begin{acknowledgments}
We would like to thank Lei Ying for helpful discussions, and Xi-Wang Luo, Pan Gao for their assistance with the calculation. This work was funded by National Natural Science Foundation of China (No. 12474366, Grants No. 11974334, and No. 11774332), and Quantum Science and Technology-National Science and Technology Major Project (Grant No. 2021ZD0301200 and No. 2021ZD0301900). We also acknowledge support from CAS Project for Young Scientists in Basic Research (Grant No. YSBR-049).
\end{acknowledgments}

\appendix
\makeatletter
\@removefromreset{equation}{section}
\@removefromreset{figure}{section}
\makeatother

\setcounter{figure}{0}
\setcounter{equation}{0}

\makeatletter
\renewcommand{\thefigure}{A\arabic{figure}}
\renewcommand{\theequation}{A\arabic{equation}}
\renewcommand{\theHfigure}{A.\arabic{figure}} 
\renewcommand{\theHequation}{A.\arabic{equation}}
\makeatother

\section{Definition and Properties of Cross Coherence Purity}

\begin{figure}
	\centering
	\includegraphics[width=8.5cm]{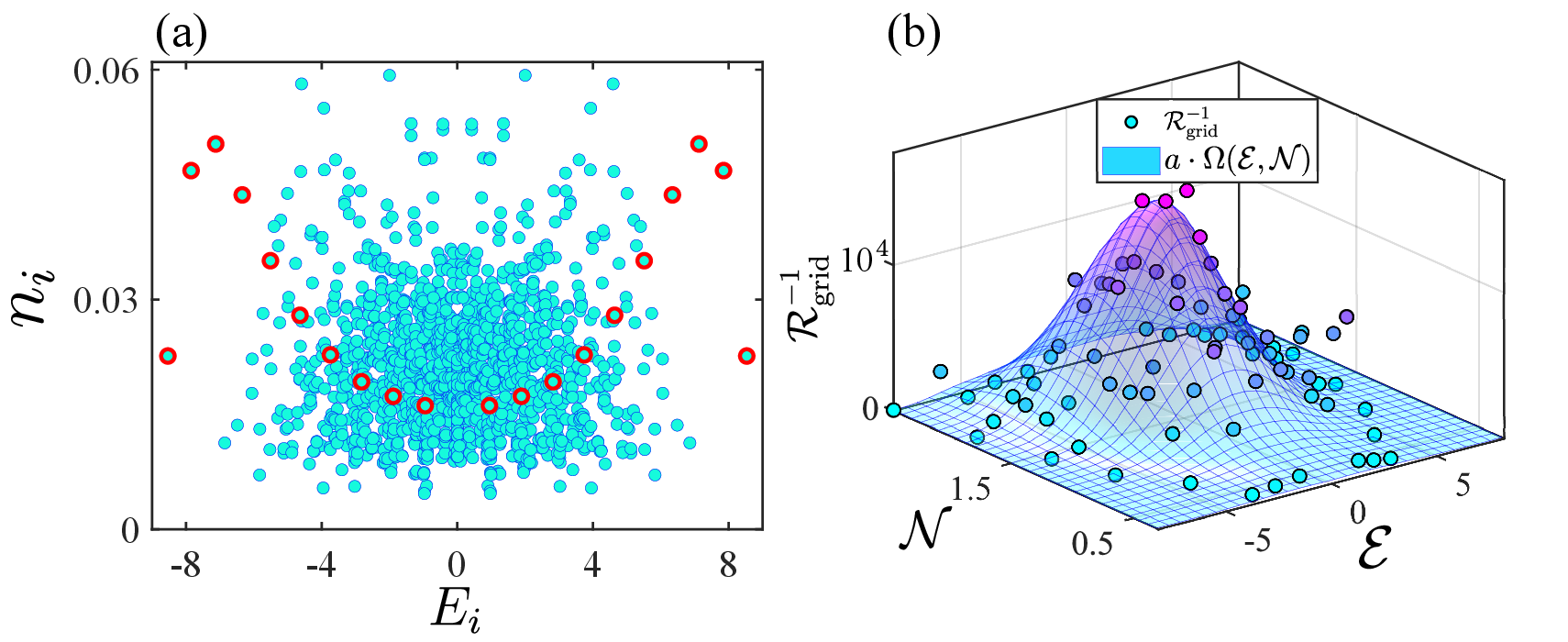}
	\caption{Plot (a) shows the numerator $n_i$ of $r_i^{\pm}$ for all eigenstates, where scar states are highlighted by red circles. 
It is evident that $n_i$ does not distinguish scar eigenstates from thermal ones.
Plot (b) displays the distribution of the off-diagonal matrix elements of the local observable $\hat{R}_k$ in the ${\cal E}$--${\cal N}$ plane. 
We collect all off-diagonal elements with energy differences in the window $0.4 < \omega < 1.6$ and define $\mathcal{R}_{i,i'} = |\langle E_i | \hat{R}_k | E_{i'} \rangle|^2$.
To suppress the effect of the pseudo-random factors, the ${\cal E}$--${\cal N}$ plane is divided into bins and $\mathcal{R}_{i,i'}$ is averaged within each bins, yielding $\mathcal{R}_{\mathrm{grid}}$.
We then plot $\mathcal{R}_{\mathrm{grid}}^{-1}$ over the ${\cal E}$--${\cal N}$ plane and fit it with $a \cdot \Omega({\cal E},{\cal N})$.
The agreement confirms that the off-diagonal matrix elements of $\hat{R}_k$ obey the modified ETH, scaling inversely with the square root of the density of states.}
\label{fig: Appendix_SGA}
\end{figure}

As defined in the main text, the Cross Coherence Purity (CCP) is a quantity constructed solely from two many-body eigenstates. It is defined as the squared Schatten-$p=2$ norm (Hilbert--Schmidt norm) of the reduced cross-density matrix $\rho_A^{i,i'} = \mathrm{Tr}_{\bar{A}}\!\left( |E_i\rangle\langle E_{i'}| \right)$, namely, $\mathcal{T}_{i,i'} \equiv \mathrm{Tr}\!\left( \rho_A^{i,i'\dagger} \rho_A^{i,i'} \right).$
Since $\rho_A^{i,i'\dagger} = \rho_A^{i',i}$, the CCP is symmetric under the exchange of its indices: $\mathcal{T}_{i,i'} = \mathcal{T}_{i',i}$.
When the two indices coincide, $i=i'$, the CCP reduces to $\mathcal{T}_{i,i} = \mathrm{Tr}\!\left[(\rho_A^{i})^2\right]$, with $\rho_A^{i} = \mathrm{Tr}_{\bar{A}}\!\left( |E_i\rangle\langle E_i| \right)$.
This is precisely the entanglement purity of the eigenstate $|E_i\rangle$ on subsystem $A$, or equivalently the second R\'enyi entropy~\cite{RenyiEntropy}. This limiting case motivates the terminology `purity' in the definition of CCP.

We now discuss how the CCP controls the scaling of off-diagonal matrix elements of local observables in the energy eigenbasis.
Consider a local operator $\hat{O}$ supported on subsystem $A$, with spectral decomposition $\hat{O} = \sum_{l} O_l |l\rangle\langle l|$.
Its off-diagonal matrix elements between two energy eigenstates read$\langle E_i | \hat{O} | E_{i'} \rangle= \mathrm{Tr}_{A}\!( \rho_A^{i',i} \hat{O})$, where $\rho_A^{i',i}$ is the reduced cross-density matrix defined above.
In the eigenbasis $\{|l\rangle\}$ of $\hat{O}$, this expression becomes
$\langle E_i | \hat{O} | E_{i'} \rangle
= \sum_{l} O_l \langle l | \rho_A^{i',i} | l \rangle$.
This allows us to derive an upper bound on the magnitude of the off-diagonal matrix element:
\begin{equation}
\begin{aligned}
|\langle E_i | \hat{O} | E_{i'} \rangle|
&\leq \max_l |O_l| \sum_l |\langle l | \rho_A^{i',i} | l \rangle| \\
&\leq \max_l |O_l| \sum_l
\sqrt{
\langle l | \rho_A^{i',i\dagger} | l \rangle
\langle l | \rho_A^{i',i} | l \rangle
} \\
&\leq \max_l |O_l| \sqrt{\mathcal{A}}
\sqrt{
\sum_l
\langle l | \rho_A^{i',i\dagger} | l \rangle
\langle l | \rho_A^{i',i} | l \rangle
} \\
&\leq \max_l |O_l| \sqrt{\mathcal{A}}
\sqrt{
\sum_{l,k}
\langle l | \rho_A^{i',i\dagger} | k \rangle
\langle k | \rho_A^{i',i} | l \rangle
} \\
&= \max_l |O_l| \sqrt{\mathcal{A}}
\sqrt{
\mathrm{Tr}\!\left( \rho_A^{i,i'\dagger} \rho_A^{i,i'} \right)
}.
\end{aligned}
\label{App_limitingNonDiagonal}
\end{equation}
Here $\mathcal{A}$ denotes the Hilbert-space dimension of subsystem $A$.
Eq.(\ref{App_limitingNonDiagonal}) follows from successive applications of the triangle inequality, the Cauchy-Schwarz inequality, and the fact that the squared diagonal elements of a matrix are bounded by its Hilbert-Schmidt norm. As a result, the magnitude of off-diagonal matrix elements of any local observable is bounded by the square root of the CCP, up to a subsystem-size-dependent prefactor. 
Since the subsystem $A$ corresponds to a local region, its Hilbert-space dimension $\mathcal{A}$ remains finite and small. Consequently, the bound is well controlled for local observables.

\section{The Approximate Spectrum-Generating Algebra for the QMBS Model}

In Sec.~V of the main text, we introduced an approximate SGA for the constrained spin model studied in this work.
This algebraic structure was originally identified in \cite{QMBS1Dspin} as a key ingredient underlying the quasi-periodic dynamics of scar states.
Here we briefly review the construction of this SGA and provide explicit definitions of the relevant operators.

We start by recalling Eq.~(\ref{SGA1}) in the main text and introduce the ladder
operators $Q^{\pm}$ as
\begin{equation}
  Q^{\pm}=\frac{Q^{y}\pm i Q^{z}}{\sqrt{2}},
  \label{QpmDefine}
\end{equation}
where $Q^{z}=\sum_{k=1}^{D}s_{k}^{z}$ is the collective spin operator along the $z$ direction, and $Q^{y}$ is obtained by rotating the constrained Hamiltonian $H$ by $\pi/2$ about the $z$ axis in spin space: $Q^y =e^{-i\frac{\pi}{2}Q^z} H e^{i\frac{\pi}{2}Q^z}$, and can be written as
\begin{equation}
\begin{aligned}
  Q^{y}= \sum_{k=1}^{D}s_{k}^{y}- i\sqrt{j}\,|x\rangle\langle y'|_{k,k+1}+ i\sqrt{j}\,|y'\rangle\langle x|_{k,k+1},
\end{aligned}
\label{Qy}
\end{equation}
where $|y'\rangle_{k,k+1} =\frac{1}{\sqrt{2}} \bigl(|j,-j+1\rangle_{k,k+1}-|j-1,-j\rangle_{k,k+1}\bigr)$.
It is worth noticing that $Q^y$ can also be viewed as the collective spin operator $\sum_{k=1}^{D}s_k^y$ constrained by the same blockades defined in Sec. II.
From these definitions, one finds the following commutation relations with the Hamiltonian: $[H,Q^{z}] = -i Q^{y}$ and $[H,Q^{y}] = i Q^{z} + \sqrt{2}\,\hat{R}$.
where the operator $\hat R$ is given by
\begin{equation}
\begin{aligned}
  \hat R &= -i\frac{j}{\sqrt{2}} \sum_{k=1}^{D} \bigl( |y'\rangle\langle y|_{k,k+1} + |y\rangle\langle y'|_{k,k+1} \bigr) \\
  & = i\frac{j}{\sqrt{2}} \sum_{k=1}^{D} \hat R_k,
\end{aligned}
\label{Rdef}
\end{equation}
with $\hat{R}_k = |j-1,-j\rangle\langle j-1,-j|_{k,k+1} - |j,-j+1\rangle\langle j,-j+1|_{k,k+1}$ as defined in the main text. 
Combining the above commutation relations, we obtain:
\begin{equation}
\begin{aligned}
  [ H , Q^{\pm} ]  &= \frac{1}{\sqrt{2}} [ H,Q^{y}\pm i Q^{z} ] \\
  &= \frac{1}{\sqrt{2}} \bigl(i Q^{z} + \sqrt{2}\hat R \pm Q^{y} \bigr) \\
  &= \pm Q^{\pm} + \hat R,
\end{aligned}
\label{APP_SGA}
\end{equation}
which reproduces the approximate SGA relation given in Eq.~(\ref{SGA1}) of the main text.

\section{Properties of the Off-Diagonal Matrix Elements of the Local Operator $\hat{R}_k$}

In the last section of the main text, we applied the modified ETH framework to the off-diagonal matrix elements of the local operator $\hat{R}_k$ and used it to construct the probe defined in Eq.~(\ref{fit_d}) for diagnosing the validity of the SGA. 
Based on this analysis, we reached the central conclusion that, within the revised ETH, the emergence of an approximately equally spaced energy structure among scar eigenstates is naturally allowed in regions of low DOS.
In this appendix, we provide a more detailed discussion of the SGA structure and the properties of the local operator $R_k$ leading to this conclusion.
We start from Eq.~(\ref{SGA2}): $H |E_i^{\pm} \rangle = (E_i \pm 1) |E_i^{\pm} \rangle + \sqrt{r_i^{\pm}}\, |r\rangle$ ,
where the quantities $r_i^{\pm}$ and $|r\rangle$ directly writes:
\begin{equation}\label{def_r_states_r}
r_i^{\pm}= \frac{\langle E_i | \hat{R}^{\dagger} \hat{R} | E_i\rangle} {\langle E_i | Q^{\mp} Q^{\pm} | E_i \rangle}, \quad |r\rangle = \frac{\hat{R}}{\sqrt{\langle E_i | \hat{R}^{\dagger} \hat{R} | E_i\rangle}}| E_i \rangle .
\end{equation}
On the other hand, from Eq.~(\ref{SGA1}) one obtains: $ Q^{\pm} |E_i\rangle=1/(H-(E_i\pm 1)) \cdot \hat{R} |E_i\rangle $.
Substituting this expression into Eq.~(\ref{SGA2}) and inserting the identity $\mathbb{I}=\sum_{i'=1}^{\mathcal D} |E_{i'}\rangle\langle E_{i'}|$ between
$\hat{R}^{\dagger}$ and $\hat{R}$ as well as between $Q^{\mp}$ and $Q^{\pm}$, the quantity $r_i^{\pm}$ can be rewritten as Eq.(\ref{ReRefine_r})
\begin{equation}
\begin{aligned}
r_i^{\pm} &= \frac{n_i}{d_i^{\pm}}, \quad 
n_i  = \sum_{i'} \big| \langle E_{i'} | \hat{R}_k | E_i \rangle \big|^2 , \\ d_i^{\pm} &= \sum_{i'} \frac{ \big| \langle E_{i'} | \hat{R}_k | E_i \rangle \big|^2}{(\omega \mp 1)^2},
\end{aligned}
\label{App_ReRefine_r}
\end{equation}
The system under consideration possesses a discrete symmetry consisting of a site-inversion ($k \to D-k+1$) followed by a global spin flip along the $z$ axis ($|m\rangle_k \to |-m\rangle_k$). We denote this combined operation by $\mathcal I_{\mathrm{SS}}$, which satisfies $[H,\mathcal I_{\mathrm{SS}}]=0$. 
Under this transformation, the local operator $\hat{R}_k$ transforms as $\hat{R}_k \;\xrightarrow{\;\mathcal I_{\mathrm{SS}}\;}\; - \hat{R}_{D-k}$.
As a consequence, within the translation-invariant subspace one has $\langle E_i | \hat{R}_k | E_i \rangle =-\langle E_i | \hat{R}_k | E_i \rangle= 0$.
Therefore, the diagonal terms $i'=i$ do not contribute in Eq.~(\ref{App_ReRefine_r}), and the summations can be restricted to $i' \neq i$, leading to Eq.~(\ref{ReRefine_r}) in the main text.
In Sec.~V, numerical results demonstrate that $r_i^{\pm}$ takes significantly smaller values for scar eigenstates than for nearby thermal eigenstates. Moreover, this distinction is shown to originate from the denominator $d_i^{\pm}$ rather than the numerator $n_i$. Indeed, as illustrated in Fig.~\ref{fig: Appendix_SGA} (a),
the quantity $n_i = \sum_{i' \neq i} \big| \langle E_{i'} | \hat{R}_k | E_i \rangle \big|^2$ does not exhibit a clear separation between scar and thermal states.

To further analyze the denominator $d_i^{\pm}$, we invoke the modified ETH, according to which the off-diagonal matrix elements of the local operator $\hat{R}_k$ scale inversely with the squared DOS $\Omega(\mathcal E,\mathcal N)$. 
As shown numerically in Fig.~\ref{fig: Appendix_SGA} (b), the inverse squared magnitude of these matrix elements is indeed proportional to $\Omega(\mathcal E,\mathcal N)$.
Since the dominant contribution to $d_i^{\pm}$ comes from matrix elements with energy differences $\omega$ close to unity, only the window $0.4 < \omega < 1.6$ is used in the statistical analysis and estimation. This observation underlies the higher $d_i^{\pm}$ for scar eigenstates and ultimately explains the emergence of their approximate SGA structure within the revised ETH framework.

\bibliography{ref_note28new}

\end{document}